\global\def\draftcontrol{0}

   \def\versionno{ boson stars in string theory}
\catcode`\@=11

\expandafter\ifx\csname draftcontrol\endcsname\relax\global\def\draftcontrol{0}
\fi

{\count255=\time\divide\count255 by 60
\xdef\hourmin{\number\count255}
\multiply\count255 by-60\advance\count255 by\time
\xdef\hourmin{\hourmin:\ifnum\count255<10 0\fi\the\count255}}
\def\draftdate{\number\month/\number\day/\number\year\ \ \ \hourmin }

\newcommand\makepapertitle{\par
  \begingroup
    \renewcommand\thefootnote{\@fnsymbol\c@footnote}%
    \def\@makefnmark{\rlap{\@textsuperscript{\normalfont\@thefnmark}}}%
    \long\def\@makefntext##1{\parindent 1em\noindent
            \hb@xt@1.8em{%
                \hss\@textsuperscript{\normalfont\@thefnmark}}##1}%
     \newpage
     \global\@topnum\z@   % Prevents figures from going at top of page.
     \@makepapertitle
     \thispagestyle{empty}\@thanks
  \endgroup
  \setcounter{footnote}{0}%
  \global\let\thanks\relax
  \global\let\makepapertitle\relax
  \global\let\@makepapertitle\relax
  \global\let\@thanks\@empty
  \global\let\@author\@empty
  \global\let\@date\@empty
  \global\let\@title\@empty
  \global\let\title\relax
  \global\let\author\relax
  \global\let\date\relax
  \global\let\and\relax
  \def\version{\let\version\@version\@gobble}
}
\def\@makepapertitle{%
  \newpage
   \ifnum\draftcontrol=1 {}
   \version\versionno
   \vskip 3em%
   \else
   \hfill\hbox to 3cm {\parbox{4cm}{\@pubnum}\hss}%
   \vskip 3em%
   \fi
   \begin{center}%
   \let \footnote \thanks
     {\LARGE {\@title}}%
     \vskip 1.5em%
     {\normalsize%\large
       \lineskip .5em%
       \begin{tabular}[t]{c}%
         \@author
       \end{tabular}\par}%
     \vskip 1.5em%
     {\@bstract}%
     \end{center}%
     \vskip 1.5em
     \@date%
   \par
}

\gdef\@pubnum{}
\def\pubnum#1{%
  \gdef\@pubnum{#1}}

\gdef\@bstract{}
\def\Abstract#1{%
  \gdef\@bstract{%
   \parbox{\textwidth-0pc}{%
   \centerline{\bf Abstract}\penalty1000%
\kern.2cm%
\noindent%\abstractfont \baselineskip=12pt
\renewcommand\baselinestretch{1.0}%
{#1}}}
}

\def\ps@paper{\let\@mkboth\@gobbletwo%
     \ifnum\draftcontrol=1
    \def\@oddfoot{\hbox to \textwidth{\tiny \versionno \hfil\tiny\draftdate}%
    \hskip -\textwidth \hbox to \textwidth{\hfil\rm\thepage\hfil}}%
     \else\def\@oddfoot{\hbox to \textwidth{\hfil\rm\thepage\hfil}}
     \fi
     \let\@evenfoot\@oddfoot
}
\def\body{\clearpage
          \pagestyle{paper}
    }
\def\@version#1{\ifnum\draftcontrol=1
\typeout{}\typeout{#1}\typeout{}
\vskip3mm\centerline{\hbox{\fbox{\normalsize{\tt DRAFT -- #1 -- }
                   {\draftdate}}}}\vskip3mm
\fi}
\let\version\@version
\long\def\eqlabel#1{\ifnum\draftcontrol=1
                    \tag@false  % there are some problems with multline without this
                    \tag*{(\theequation) \hbox to -0.2cm{\hspace{0cm}\small{#1}\hss}}
                    \refstepcounter{equation}
                    \edef\@currentlabel{\theequation}
                    \ltx@label{#1}          % use old LaTeX \label instead of new definition
                    \else
                    \label{#1}
                    \fi
                    }
\let\st@bibitem\@bibitem
\let\st@lbibitem\@lbibitem
\ifnum\draftcontrol=1
  \def\@bibitem#1{%
    \st@bibitem{#1}\a@@label{#1}\ignorespaces}
  \def\@lbibitem[#1]#2{%
    \st@lbibitem[#1]{#2}\a@@label{#2}\ignorespaces}
  \def\a@@label#1{%
    \gdef\a@lab{\smash{\normalfont\small#1}}
    \ifvmode
      \if@inlabel
        \global\setbox\@labels\hbox{%
          \llap{\a@lab\let\a@lab\relax
                \kern\@totalleftmargin\kern\marginparsep}%
          \box\@labels}%
      \fi
    \fi}
\fi
\documentclass[12pt,letterpaper]{article}
\usepackage{amsmath,amssymb,array,calc,epsfig,rotating,bm}
\usepackage[sort]{cite}
\usepackage{graphicx}
\usepackage{psfrag,verbatim}

\ifnum\draftcontrol=1
\fi
\renewcommand\baselinestretch{1.25}
\renewcommand\section{\@startsection {section}{1}{\z@}%
                                   {-3.5ex \@plus -1ex \@minus -.2ex}%
                                   {2.3ex \@plus.2ex}%
                                   {\normalfont\large\bfseries}}
\renewcommand\subsection{\@startsection{subsection}{2}{\z@}%
                                   {-3.25ex\@plus -1ex \@minus -.2ex}%
                                   {1.5ex \@plus .2ex}%
                                   {\normalfont\normalsize\bfseries}}
\renewcommand\subsubsection{\@startsection{subsubsection}{3}{\z@}%
                                   {-3.25ex\@plus -1ex \@minus -.2ex}%
                                   {1.5ex \@plus .2ex}%
                                   {\normalfont\normalsize\it}}
\renewcommand\paragraph{\@startsection{paragraph}{4}{\z@}%
                                   {-3.25ex\@plus -1ex \@minus -.2ex}%
                                   {1.5ex \@plus .2ex}%
                                   {\normalfont\normalsize\bf}}

\numberwithin{equation}{section}

\def\revise#1       {\raisebox{-0em}{\rule{3pt}{1em}}%
                     \marginpar{\raisebox{.5em}{\vrule width3pt\
                     \vrule width0pt height 0pt depth0.5em
                     \hbox to 0cm{\hspace{0cm}{%
                     \parbox[t]{4em}{\raggedright\footnotesize{#1}}}\hss}}}}

\newcommand\nxt[1]  {\\\fnxt#1}
\newcommand{\ie}{{\it i.e.,}\ }

\def\calm         {{\cal M}}
\def\caln         {{\cal N}}
\def\calo         {{\cal O}}

\def\calv         {{\cal V}}
\def\calw         {{\cal W}}

\def\del          {\partial}

\def\sqr#1#2{{\vcenter{\vbox{\hrule height.#2pt
 \hbox{\vrule width.#2pt height#1pt \kern#1pt
 \vrule width.#2pt}\hrule height.#2pt}}}}

\def\a{\alpha}

\def\w{\omega}

\def\dd{\delta}
\def\e{\epsilon}

\def\aa1{\phi}
\def\cc1{\psi}

\def\l{\lambda}

\def\hM{\hat{M}}
\def\hQ{\hat{Q}}

\def\vol{{\rm vol}}

\catcode`\@=12

\begin{document}

%%%
%%%%%% text starts here
%%%%%%%%%

\title{\bf AdS boson stars in string theory}

\date{October 27, 2015}
%\date\today

\author{
Alex Buchel \\[0.4cm]
\it Department of Applied Mathematics, Department of Physics and Astronomy, \\
\it University of Western Ontario\\
\it London, Ontario N6A 5B7, Canada;\\
\it Perimeter Institute for Theoretical Physics\\
\it Waterloo, Ontario N2J 2W9, Canada
}

\Abstract{Boson stars are stationary soliton-like gravitational configurations
supported by a complex scalar field charged under the global $U(1)$
symmetry.  We discuss properties of boson stars in type IIB
supergravity approximation to string theory. A notable difference is
that in supergravity models the global symmetry of the complex scalar
field is gauged.  We focus on global asymptotically $AdS_5$
space-time, where the boson stars are expected to represent stable
low-energy non-thermal excitations in holographically dual quiver
conformal gauge theory.
}

\makepapertitle

\body

\version\versionno
\tableofcontents

\section{Introduction}\label{intro}
Consider an effective gravitational action containing Einstein-Hilbert term and a complex scalar field $\Psi$
with an exact global $U(1)$ symmetry,
\begin{equation}
S_{d+1}=\frac{1}{16\pi G_{d+1}}\int_{\calm_{d+1}} dx^{d+1}\sqrt{-g}\  \left(R -\frac 12 \del\Psi\del\bar\Psi-V(\Psi\bar\Psi)\right)\,,
\eqlabel{toybs}
\end{equation}
where $V$ is a scalar field potential, and $\Psi\to e^{i\a} \Psi $ under the symmetry transformations.
The latter symmetry allows for an interesting class of gravitational soliton-like solutions
where the scalar field has a harmonic time dependence,  with a nonetheless static stress-energy tensor, 
supporting the curved background metric on $\calm_{d+1}$. Such configurations, named {\it boson stars}, 
were extensively studied in asymptotically Minkowski space-time \cite{Liebling:2012fv}
and in asymptotically AdS space-time beginning with the work \cite{Astefanesei:2003qy}\footnote{
The importance of boson stars for the AdS instability problem \cite{Bizon:2011gg} was emphasized in 
\cite{Buchel:2013uba}.}.
 
Boson stars are expected to play an important role in holographic gauge theory/string theory correspondence 
\cite{Aharony:1999ti}. Specifically, thermal states in microcanonical ensemble of strongly coupled four-dimensional 
conformal gauge theories (CFT) on $S^{3}$ are represented by Schwarzschild (global) $AdS_5$ black holes (BH), smeared over the compact manifold 
$\calv_5$ encoding the global symmetries of the CFT. This picture is correct for sufficiently large black holes in AdS (sufficiently energetic 
states in the CFT) \cite{Horowitz:1999uv}: as the BH gets smaller, it ultimately develops a Gregory-Laflamme instability \cite{Gregory:1993vy}  
associated with its localization on $\calv_5$ \cite{Hubeny:2002xn,Dias:2015pda,Buchel:2015gxa,Buchel:2015pla}. The localization 
process corresponds to the dynamical spontaneous global symmetry breaking of the dual CFT. This suggests that
supersymmetric conformal gauge theories on $S^3$ can not have stable equilibrium states with unbroken $R$-symmetry below certain 
energy threshold. It was proposed in \cite{Buchel:2015sma} that stable low-energy stationary states in $R$-symmetry singlet sector 
of  conformal gauge theories are represented by boson stars in the gravitational dual.        

Until now, the boson stars were discussed in the context of phenomenological gravity-scalar models. In this paper we 
construct boson stars in type IIB supergravity backgrounds  holographically dual to strongly coupled 
quiver conformal gauge theories introduced in \cite{Gubser:2009qm} (GHPT), 
and initiate analysis of their properties. Unlike the boson stars discussed earlier, 
boson stars arising in holography have local $U(1)$ symmetry. 

The rest of the paper is organized as follows. In the next section we review the correspondence 
between strongly coupled $\caln=1$ superconformal gauge theories and holographically dual five-dimensional 
consistent truncations of type IIB supergravity  introduced in \cite{Gubser:2009qm}. We discuss in details 
the thermal equilibrium states in the theory with and without the condensate of the complex 
scalar field\footnote{In the former case we generalize the "holographic superconductor''
transition for CFT on $R^{3,1}$ in \cite{Gubser:2009qm} to curved space-time, $R\times S^3$.}. 
In section \ref{toy} we discuss boson stars in a toy model obtained from  GHPT effective action  
removing the bulk gauge field (turning off the gauge coupling of the complex scalar field) 
and approximating the scalar potential with the appropriate mass term. We study boson stars in GHPT 
effective action in section \ref{full}. Finally, we conclude in section \ref{conclude}.

\section{Quiver $\caln=1$ supersymmetric CFT holography}\label{holography}

In this section we discuss the holographic correspondence between a large class of strongly coupled 
$\caln=1$ superconformal quiver gauge theories and GHPT consistent truncation of type IIB supergravity 
on five-dimensional Sasaki-Einstein manifolds. We then study thermal equilibrium states within this 
consistent truncation.

\subsection{GHPT effective action}

At weak 't Hooft coupling, the gauge theory living on a world-volume of a large number $N$ of D3-branes places at the tip of a three complex dimensional 
Calabi-Yau cone $X$ in type IIB supergravity is $\caln=1$ superconformal (SCFT) 
quiver gauge theory with $SU(N)$ gauge groups and a certain superpotential \cite{Kehagias:1998gn,Klebanov:1998hh,Acharya:1998db,Morrison:1998cs}.  
In the planar limit and for large 't Hooft coupling the theory is best described by type IIB supergravity on 
$AdS_5\times Y_5$, where the Einstein-Sasaki manifold $Y_5$ is a level surface of $X$.  Following \cite{Gubser:2009qm}
we consider examples where $Y_5$ is expressible as a $U(1)$ fibration over a compact K\"ahler-Einstein base. In this case the symmetry of the 
fiber geometrizes the $R$-symmetry of the SCFT.  

Depending of which aspects of the dual gauge theories one wishes to study, one uses different consistent truncations of type IIB 
supergravity on $AdS_5\times Y_5$. 
For example, to study the hydrodynamic transport in the theory \cite{Buchel:2008ae} it is enough to consistently truncate to a 
five-dimensional gravity sector with a negative cosmological constant. Study of states of the theory with $R$-symmetry 
charge requires gauging the $U(1)$ fiber isometry of $Y_5$ \cite{Buchel:2006gb}.  Finally, to study boson stars in these SCFT 
one needs to enlarge the consistent truncation \cite{Buchel:2006gb} to include a complex scalar field $\Psi$, charged under the 
$R$-symmetry. In a special case when $\Psi$ is dual to a chiral primary operator $\calo_\Delta$ of a CFT with a scaling dimension 
$\Delta=3$ such a truncation was constructed in   \cite{Gubser:2009qm}. Explicitly,  GHPT effective action takes 
form
\begin{equation}
\begin{split}
S_{GHPT}=&\frac{1}{16\pi G_5}\int_{\calm_5} dx^5\sqrt{-g} \biggl(R-\frac{L^2}{3} F_{\mu\nu}F^{\mu\nu}+\left(\frac{2L}{3}\right)^3
\frac 14 \e^{\lambda\mu\nu\sigma\rho} F_{\lambda\mu}F_{\nu\sigma}A_\rho\\
&-\frac 12 \biggl[(\del\eta)^2+\sinh^2\eta\ (\del_\mu\theta -q\ A_\mu)^2 -\frac{6}{L^2}\ \cosh^2\frac\eta2\ (5-\cosh\eta)\biggr]
\biggr)\,,
\end{split}
\eqlabel{ghptaction}
\end{equation}
where $\eta$ and $\theta$ and the modulus and the phase of the complex scalar $\Psi\equiv \eta e^{i\theta}$, 
the five-dimensional Newtons constant is 
\begin{equation}
G_5=\frac{G_{10}}{{\rm vol}(Y_5)}\,.
\end{equation} 
The $R$-charge of $\Psi$, dual to the chiral primary $\calo_3$, 
\begin{equation}
|\ R[\Psi]\ | = q=\frac 23 \Delta = 2\,,
\eqlabel{charge}
\end{equation}
 is fixed by the superconformal algebra\footnote{We use the standard 
normalization where the $R$-charge of the SCFT superpotential $\calw$ is $R[\calw]=2$.}. In what follows we set the asymptotic 
$AdS_5$ radius $L=1$. The ten-dimensional uplift of \eqref{ghptaction} can be found in \cite{Gubser:2009qm}.

\subsection{Thermal state without scalar condensates in microcanonical ensemble}

Consider thermal states in GHPT gauge theory plasma without the scalar condensate, \ie 
$\eta=0$. These states are represented in a holographic dual by an electrically charged (RN) black
hole. We discuss RN black hole solution in some details mainly to set up our conventions. 

We use the five-dimensional background metric as 
\begin{equation}
ds_5^2= \frac 1y \biggl[-a e^{-2\dd} (dt)^2 +\frac{(dy)^2}{4y(1-y)a}+(1-y) (d\Omega_3)^2\biggr]\,,
\eqlabel{metric5d}
\end{equation}
where $(d\Omega_3)^2$ is a metric on a round $S^3$ of unit radius, and $\{a,\dd\}$ are the warp factors depending on the 
radial coordinate $y$ varying from the AdS boundary $y=0$ to the location of the 
regular  Schwarzschild horizon at $y=y_0 < 1$, $y\in[0,y_0]$. The background further has a nontrivial bulk gauge potential 
\begin{equation}
A=\phi(y)\ d(t)\,.
\eqlabel{bulkgauge}
\end{equation}
Solving the equations of motion we find 
\begin{equation}
\begin{split}
&\phi=\frac{\mu( y_0-y)}{y_0(1-y)}\,,\qquad \dd=0\,,\\
&a=\frac{(y_0-y) (y^2(4 \mu^2+9) (y_0-1)-18 y y_0+9 y+9 y_0)}{9(1-y)^2 y_0^2}\,.
\end{split}
\eqlabel{rnsolve}
\end{equation}
Following the AdS/CFT dictionary, the normalizable and the non-normalizable coefficients of various gravitational 
modes encode the CFT data. From a general AdS boundary expansion,
\begin{equation}
\begin{split}
\phi=\mu+\phi^b_1\ y+\calo(y^2)\,,\qquad a=1+a^b_2\ y^2 +\calo(y^3) \,.
\end{split}
\eqlabel{uv}
\end{equation}
$\mu$ is the RN black hole chemical potential, and the normalizable components 
\begin{equation}
\phi^b_1=-\frac{\mu}{y_0}+\mu\,,\qquad a_2^b=\frac{(y_0-1)(4\mu^2 y_0+9)}{9y_0^2}\,, 
\eqlabel{rndata}
\end{equation}
determine the charge $Q$  and the mass $M$ of RN black hole as
\begin{equation}
M= \frac 34\ \frac{\vol(S^3)}{16\pi G_5}\ (1-4a_2^b)\,,\qquad Q=\frac{\vol(S^3)}{2\pi G_5}\ \left(-\frac 13 \phi^b_1\right)\,.
\eqlabel{mq}
\end{equation}
Additionally, we can compute the temperature $T$, the entropy $S$ and the grand potential $\Omega$ of  RN black hole as   
\begin{equation}
\begin{split}
&(\pi T)^2=\frac{((4\mu^2+9)y_0-18)^2}{324y_0(1-y_0)}\,,\qquad ST =\frac{\vol(S^3)}{16\pi G_5}\ \frac{2(1-y_0)(18-y_0(4\mu^2+9))}{9y_0^2}\,, \\
&\Omega=\frac{\vol(S^3)}{16\pi G_5}\ \ \biggl(\frac{(5y_0-2)(2-y_0)}{4y_0^2}-\frac{4(1-y_0)}{9 y_0}\ \mu^2\biggr)\,.
\end{split}
\eqlabel{rnrest}
\end{equation}
Notice that the basic thermodynamic relations are satisfied:
\begin{equation}
\Omega=M-ST -\mu Q\,,\qquad  d(\Omega)=-S\ d(T)- Q\ d(\mu)\,.
\eqlabel{thermorel}
\end{equation}
The vacuum of a SCFT is represent by global $AdS_5$ solution, $y_0=1$ and $\mu=0$, in which case 
\begin{equation}
M_{AdS_5}=  \frac 34\ \frac{\vol(S^3)}{16\pi G_5}
\eqlabel{casimir}
\end{equation}
is the Casimir energy of the CFT ground state. 
In what follows we find it convenient to introduce the reduced mass $\hM$ and the charge $\hQ$ as follows
\begin{equation}
\hM=\frac{M}{M_{AdS_5}}-1\,,\qquad \hQ= \frac{2\pi G_5}{\vol(S^3)}\ Q\,.
\eqlabel{mqred}
\end{equation}  

In a microcanonical ensemble, we keep the mass and the charge of a BH fixed, in which case 
we find 
\begin{equation}
\hM(\hQ,y_0)=\frac{4(1-y_0)}{y_0^2}+\frac{16y_0}{1-y_0}\ \hQ^2\,.
\eqlabel{mqcurve}
\end{equation}
Notice that for a fixed $\hQ$, 
\begin{equation}
\hM(\hQ,y_0)\ge \hM_{min}(\hQ)=16\ \hQ+16\ \hQ^2-32\ \hQ^3+\calo(\hQ^4)\,,
\eqlabel{mmin}
\end{equation}
for small $\hQ$; for general $\hQ$, $\hM_{min}(\hQ)$ can be computed numerically.

\begin{figure}[t]
\begin{center}
\psfrag{q}{{$\hQ$}}
\psfrag{m}{{$\hM$}}
\psfrag{d}{{$\hM_{crit}-\hM_{min}$}}
\includegraphics[width=3.0in]{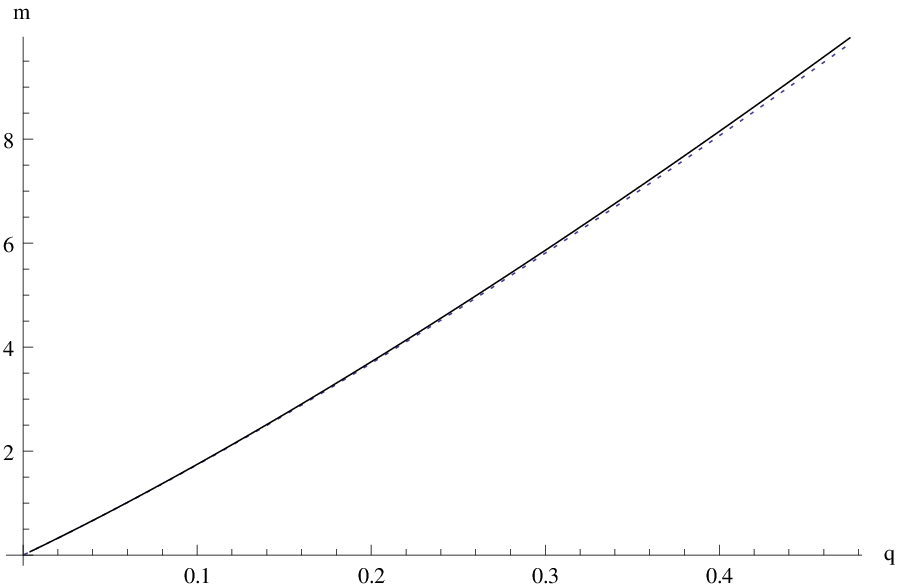}
\includegraphics[width=3.0in]{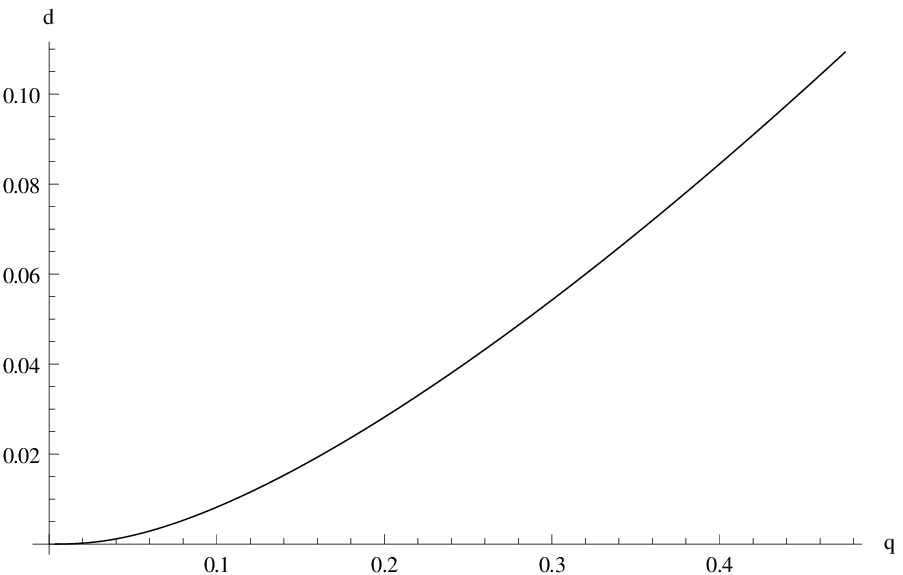}
\end{center}
  \caption{Left panel: Solid black line indicates the onset of the global $AdS_5$ RN black hole instability with respect to  
developing a condensate of a charged scalar in GHPT model. The dotted blue line is the minimal mass of the 
RN BH for a given charge $\hQ$, see \eqref{mmin}. The RN black holes in a wedge between the blue and the black lines are unstable. 
This wedge is enlarged in the right panel. } \label{figure1}
\end{figure}

In GHPT plasma, described holographically by \eqref{ghptaction}, fixed charge equilibrium states 
\eqref{mq}, \eqref{rnrest} become unstable with respect to the condensation of the complex 
scalar $\Psi$ at low-energies --- this is the "holographic superconductor'' transition of  \cite{Gubser:2009qm} for a 
plasma in $R^3$. Similar phenomenon\footnote{We found that  
GHPT ``superconducting'' transition on $R^3$ occurs at $T=0.0606766(2) \mu$, in agreement 
with the result reported in \cite{Gubser:2009qm}.} occurs when  GHPT plasma is confined on $S^3$.
To determine the onset of the instability we linearize equation for\footnote{We explicitly
factor the leading asymptotic behavior of $\eta$ in defining $\eta_1$.} $\eta(y)\equiv y^{3/2} \eta_1(y)$ 
(we can choose the gauge with $\theta=0$)
on global RN black hole background \eqref{metric5d}-\eqref{rnsolve}. For a fixed $\hQ$, this equation takes form
\begin{equation}
\begin{split}
&0=\eta_1''+\biggl(4 y^2 y_0^2 (5 y-4 y_0) \hQ^2+(y_0-1) (5 y^3 y_0-4 y^2 y_0^2-5 y^3-4 y^2 y_0
+6 y y_0^2+4 y^2\\
&-2 y_0^2)\biggr)\biggl(4 y^3 (-y_0+y) y_0^2 \hQ^2+y (y_0-y) (1-y_0) (y-1) (y y_0-y-y_0)\biggr)^{-1} \eta_1'
\\&+\biggl(48 y^3 y_0^4 (5 y-3 y_0) \hQ^4+12 y_0^2 (10 y^4 y_0^2-6 y^3 y_0^3-20 y^4 y_0-4 y^3 y_0^2+9 y^2 y_0^3
+10 y^4\\
&+18 y^3 y_0-10 y^2 y_0^2-8 y^3-2 y^2 y_0+6 y y_0^2-3 y_0^3) \hQ^2
+3 (y_0-1)^2 (y-1) (y y_0-y\\
&-y_0) (5 y^2 y_0-3 y y_0^2-5 y^2-3 y y_0+3 y_0^2+3 y)\biggr)
\biggl(64 y^5 y_0^4 (-y_0+y) \hQ^4+32 (y_0-1) (y\\
&-1) (y y_0-y-y_0) (y-y_0) y^3 y_0^2 \hQ^2
+4 (y_0-1)^2 (y-1)^2 (y y_0-y-y_0)^2 (y-y_0) y\biggr)^{-1} \eta_1\,.
\end{split}
\eqlabel{etal}
\end{equation}
Without loss of generality we can normalize $\eta_1$ at the horizon to be one.
Then, \eqref{etal} has to be solved requiring normalizability of $\eta$ (regularity of $\eta_1$) at the $AdS_5$ boundary
\begin{equation}
\eta_1=\eta_{1,0} \biggl(1-\frac{9(4\hQ^2 y_0^2-(y_0-1)^2)y}{8(y_0-1)^2}+\calo(y^2)\biggr)\,,
\eqlabel{uveta}
\end{equation}
and regularity of $\eta_1$ at the BH horizon
\begin{equation}
\eta_1=1+\frac{3(8 \hQ^2 y_0^3+(2 y_0-3) (y_0-1)^2)}{4(4 \hQ^2 y_0^3+(y_0-2) (y_0-1)^2) y_0}\ (y_0-y)+\calo((y_0-y)^2)\,.
\eqlabel{ireta}
\end{equation}
Notice that given $\hQ$, the regularity of the $\eta_1(y)$ solution to the second order ODE \eqref{etal} uniquely 
determines $\{y_0,\eta_{1,0}\}$. The obtained value of $y_0$ can then we used in \eqref{mqcurve} to compute the mass of the 
RN black hole corresponding to the onset of the $\Psi$-condensation instability. The results of this analysis are presented in 
Figure \ref{figure1}. On the left panel, 
the solid black line corresponds to the mass $\hM_{crit}(\hQ)$ of the RN black hole below which it becomes 
unstable to developing $\Psi$-condensate.   The dotted blue line represents  the minimal mass of the RN black 
hole for a given $\hQ$, see \eqref{mmin}. On the right panel the solid black line represents the difference 
 $(\hM_{crit}(\hQ)-\hM_{min}(\hQ))$.

\subsection{Thermal states with scalar condensate in microcanonical ensemble}\label{bhbh}

The end point of the RN BH instability in GHPT model, see \eqref{ghptaction},
 is a black hole with a nontrivial condensate of the complex scalar.
We refer to it as RN-$\Psi$ black hole.
Using the gauge $\theta=0$, the background ansatz \eqref{metric5d} and \eqref{bulkgauge} supplemented with 
\begin{equation}
c\equiv\cosh\eta=c(y)\,,
\eqlabel{defc}
\end{equation}   
we obtain the following equations of motion (in GHPT model $q=2$):
\begin{equation}
\begin{split}
0=&a'+\frac49 (y-1) y^2 e^{2\delta}\ (\phi')^2+\frac{(y-1) a y}{3(c^2-1)}\ (c')^2
- \frac{(c^2-1) q^2 \phi^2}{12a}\ e^{2\delta}-\frac{(c-1)(c-3)}{4y}\\
&-\frac{(y-2) (a-1)}{y (y-1)}\,,
\end{split}
\eqlabel{bh1}
\end{equation}
\begin{equation}
\begin{split}
&0=\delta'+\frac{y (y-1)}{3(c^2-1)}\ (c')^2- \frac{(c^2-1) q^2 \phi^2}{12a^2}\ e^{2\delta}\,,
\end{split}
\eqlabel{bh2}
\end{equation}
\begin{equation}
\begin{split}
0=&c''-\frac{c}{c^2-1}\  (c')^2- \frac{4(y-1)y^2}{9a}\ e^{2\delta} c' (\phi')^2
+\frac{(c-2)^2 (y-1)+a (8 y-4) -5 y+9)}{4(y-1) a y}\ c'\\
&-\frac{c q^2 \phi^2  (c^2-1)}{4y (y-1) a^2}\ e^{2\delta} 
+ \frac{3(c^2-1) (c-2)}{4a (y-1) y^2}\,,
\end{split}
\eqlabel{bh3}
\end{equation}
\begin{equation}
\begin{split}
&0=\phi''-\frac{(y-1) y}{3(c^2-1)}\ \phi' (c')^2+\frac{(c^2-1) q^2 \phi^2}{12a^2}\ e^{2\delta}\phi' 
+\frac{2}{y-1}\  \phi'+\frac{3q^2 (c^2-1) \phi}{16a (y-1) y^2}\,.
\end{split}
\eqlabel{bh4}
\end{equation}
Equations \eqref{bh1}-\eqref{bh4} must be solved with the following asymptotic expansions:
\nxt at the $AdS_5$ boundary, $y\to 0_+$,
\begin{equation}
\begin{split}
&a=1+a^b_2\ y^2+\left(\frac 49 (\phi^b_1)^2+c^b_3+a^b_2\right)\ y^3+\calo(y^4)\,,\qquad \dd=\frac 12 c^b_3\ y^3+\calo(y^4)\,,\\
&c=1+c^b_3\ y^3 +\calo(y^4)\,,\qquad \phi=\mu+\phi^b_1\ y +\phi^b_1\ y^2+\left(\phi^b_1+\frac 14 \mu c^b_3\right)\ y^3+
\calo(y^4)\,;
\end{split}
\eqlabel{bbhuv}
\end{equation}
\nxt at the regular Schwarzschild horizon, $z\equiv  y_0-y\to 0_+$,
\begin{equation}
\begin{split}
&a=\biggl(\frac{5-(c^h_0-2)^2}{4y_0}+\frac49 (y_0-1) y_0^2 (d^h_0)^2 (\phi^h_1)^2+\frac{1}{y_0 (1-y_0)}\biggr)\ z+\calo(z^2)\,,\\
&\dd=\ln d^h_0+\frac{27 (y_0-1) ((c^h_0)^2-1) (16 (d^h_0)^2 (\phi^h_1)^2 y_0^3 (1-y_0)+9 (c^h_0-2)^2)}
{y_0 (-16 (d^h_0)^2 (\phi^h_1)^2 y_0^3 (1-y_0)^2+9 ((c^h_0-2)^2-5) (y_0-1)+36)^2}\ z+\calo(z^2)\,,\\
&c=c^h_0+\frac{27 ((c^h_0)^2-1) (c^h_0-2)}{y_0 (-16 (d^h_0)^2 (\phi^h_1)^2 y_0^3 (1-y_0)^2+9 ((c^h_0-2)^2-5) (y_0-1)+36)}\ z+\calo(z^2)\,,\\
&\phi=\phi^h_1\ z+\calo(z^2)\,.
\end{split}
\eqlabel{bbhir}
\end{equation}
Notice that given the non-normalizable coefficients $\mu$ (the chemical potential) and the black hole size $0<y_0<1$,  
the asymptotics of the solution are completely characterized 
by 6 parameters
\begin{equation}
\{\phi^b_1\,,\ a^b_2\,,\ c^b_3\,,\ d_0^h\,,\ c^h_0\,,\ \phi^h_1\}\,,
\eqlabel{parameters}
\end{equation}
which is the correct number to uniquely specify the solution to a coupled 
system of 2 second-order equations \eqref{bh3}, \eqref{bh4} and 
2 first-order equations \eqref{bh1}, \eqref{bh2}: $2\times 2 +2\times 1= 6$. 
The mass and the charge of  RN-$\Psi$ black hole are given as 
in \eqref{mq} and \eqref{mqred}\footnote{We also computed the 
remaining thermodynamic quantities in the grand canonical ensemble and verified
the first law of thermodynamics \eqref{thermorel} to an accuracy 
of better than $10^{-6}$.}:
\begin{equation}
\hM=-4 a^b_2\,,\qquad \hQ=-\frac 13  \phi^b_1\,.
\eqlabel{mqhair}
\end{equation}

\begin{figure}[t]
\begin{center}
\psfrag{x}{{$y_0$}}
\psfrag{y}{{$\hM$}}
\includegraphics[width=4.0in]{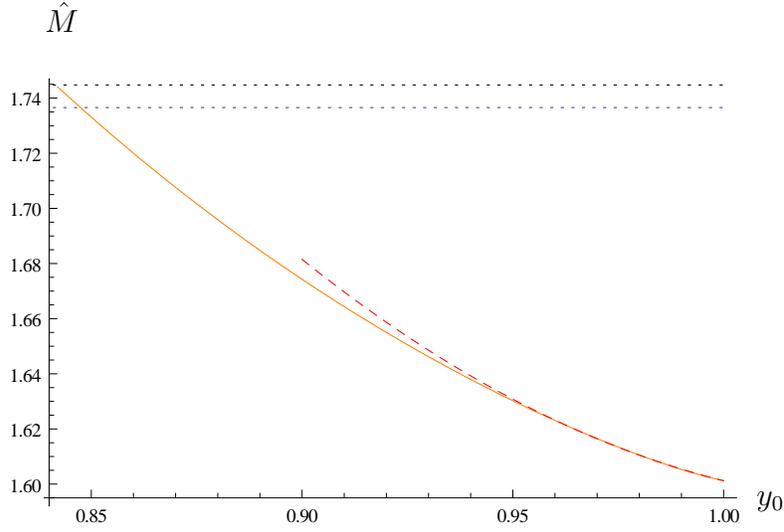}
\end{center}
  \caption{Solid orange line represents a family of RN-$\Psi$
black holes (charged black holes with nontrivial scalar hair)
for a fixed $\hQ=0.1$. The dotted black line is the largest mass of the 
unstable RN black hole with this charge $\hQ$, see \eqref{fig21}. 
The dotted blue line is the smallest mass of the RN black hole with this 
charge $\hQ$, see \eqref{fig22}. The dashed red line is the best quadratic 
fit to the mass of small RN-$\Psi$ black holes, see \eqref{redfit}.
} \label{figure2}
\end{figure}

In a microcanonical ensemble it is appropriate to keep $\hQ$ fixed and 
construct a family of RN-$\Psi$ solutions parameterized by $y_0$. As the  
area of the BH horizon is given by
\begin{equation}
A_{horizon}=\frac{(1-y_0)^{3/2}}{y_0^{3/2}}\ \vol(S^3) \,,
\eqlabel{horizon}
\end{equation}
the limit $y_0\to 0$ corresponds to a vanishingly small BH. 
We use numerical techniques of \cite{Aharony:2007vg} to construct 
families of RN-$\Psi$ black holes for different values of $\hQ$. 
A typical family  (with $\hQ=0.1$) 
is presented in Figure \ref{figure2}. The solid orange curve 
extends from $y_0=0.842$ (the RN black hole becomes 
unstable at a slightly smaller value of $y_0=0.841520144(9)$)
to $y_0=0.9999$ (the smallest black hole we studied). The dotted 
black line is the mass of the RN black hole at the onset of the 
$\Psi$-instability, 
\begin{equation}
\hM_{unstable}\bigg|_{\hQ=0.1}=1.744761(0)\,.
\eqlabel{fig21}
\end{equation}
The dotted blue line represents 
the lightest RN black hole with the  charge $\hQ=0.1$, 
\begin{equation}
\hM_{min}\bigg|_{\hQ=0.1}=1.736561(2)\,.
\eqlabel{fig22}
\end{equation}
The dashed red line is the best quadratic fit to the tail 
of the orange curve\footnote{We use the last 50 data points, 
 $y_0\in[0.959,0.9999]$.},
\begin{equation}
\hM\bigg|_{red\,, dashed}=1.601197(5) + 0.381186(8)\ (1 - y_0) + 
 4.219031(9)\ (1 - y_0)^2\,.
\eqlabel{redfit}  
\end{equation} 

We would like to conclude this section with two observations:
\begin{itemize}
\item RN-$\Psi$ black holes can have a mass smaller than that of a minimal 
mass of a RN black hole for a given charge $\hQ$;
\item in all examples we studied there is a bound on a minimal mass 
of  RN-$\Psi$ black holes  with a fixed charge, achieved when the 
corresponding black hole becomes vanishingly small; furthermore,
\begin{equation}
\min_{\hQ=const}\biggl[ \hM_{{\rm RN}-\Psi}\biggr]\ > 16 \hQ\,.
\eqlabel{minrnpsi}
\end{equation} 
\end{itemize}

\section{Boson stars in a toy model}\label{toy}

Setting $q=0$, decoupling the bulk gauge filed $A_\mu=0$, and expanding GHPT effective action 
\eqref{ghptaction} to quadratic order in $\eta$ we find
\begin{equation}
\begin{split}
S_{toy}=&\frac{1}{16\pi G_5}\int_{\calm_5} dx^5\sqrt{-g} \biggl(R+12-\frac 12 \biggl[(\del\eta)^2+\eta^2\ (\del\theta)^2 -3\eta^2\biggr]
\biggr)\,,
\end{split}
\eqlabel{toy1}
\end{equation}
which is the effective action for the phenomenological boson stars \eqref{toybs} with 
\begin{equation}
\Psi=\eta e^{i\theta}\,,\qquad V(\Psi\bar\Psi)=-12-\frac 32 \Psi\bar\Psi\,.
\eqlabel{mappingpheno}
\end{equation}
Boson star solutions are found within the metric ansatz \eqref{metric5d} supplemented with 
\begin{equation}
\eta=\eta(y)\,,\qquad  \theta=\w t \,,
\eqlabel{bsadd}
\end{equation}
where $\w$ is a constant frequency of a boson star. Equations of motion and the asymptotic data from \eqref{toy1} take form
\begin{equation}
\begin{split}
0=&a'+\frac13 a y (y-1) (\eta')^2-\frac{\eta^2 \w^2 e^{2\dd}}{12a}-\frac{\eta^2 (1-y)+4 (y-2) (a-1)}{4y (y-1)}\,,
\end{split}
\eqlabel{bstoye1}
\end{equation}
\begin{equation}
\begin{split}
0=&\dd'+\frac13 y (y-1) (\eta')^2-\frac{\w^2 \eta^2 e^{2\dd}}{12a^2}\,,
\end{split}
\eqlabel{bstoye2}
\end{equation}
\begin{equation}
\begin{split}
0=&\eta''+\frac{\eta^2 (1-y)+8 y a-4 y-4 a+8}{4a y (y-1)} \eta'-\frac{(e^{2\dd} \w^2 y+3 a) \eta}{4(y-1) y^2 a^2}\,;
\end{split}
\eqlabel{bstoye3}
\end{equation}
\nxt at the $AdS_5$ boundary, $y\to 0_+$, 
\begin{equation}
\begin{split}
&a=1+a^b_2\ y^2+\left(a^b_2+\frac12 (\eta^b_3)^2\right)\ y^3+\calo(y^4)\,,\qquad \dd=\frac14 (\eta^b_3)^2\ y^3+\calo(y^4)\,,\\
&\eta=y^{3/2}\ \biggl(\eta^b_3-\frac18 \eta^b_3 (\omega^2-9)\ y+\calo(y^2)\biggr)\,;
\end{split}
\eqlabel{bstoyuv}
\end{equation}
\nxt at the origin, $z\equiv 1-y\to 0_+$,
\begin{equation}
\begin{split}
&a=1-\frac{1}{24} (\eta^h_0)^2 ((d^h_0)^2\omega^2-3)\ z+\calo(z^2)\,,\qquad \dd=\ln d^h_0-\frac{1}{12}(\eta^h_0)^2(d^h_0)^2\w^2\ z+\calo(z^2)\,,\\
&\eta=\eta^h_0-\frac 18 \eta^h_0((d^h_0)^2\omega^2+3)\ z+\calo(z^2)\,.
\end{split}
\eqlabel{bstoyir}
\end{equation}

Effective  action \eqref{toy1} has a global $U(1)$ symmetry associated with $\theta\to\theta+\a$. The corresponding conserved current  
\begin{equation}
J^\mu=-\frac 14 \sqrt{-g} \eta^2 \del^\mu \theta\,,\qquad \del_\mu J^\mu =0\,, 
\eqlabel{bscurrent}
\end{equation}
gives a charge density 
\begin{equation}
\hQ=-\frac 14\int_0^1 dy \sqrt{-g} \eta^2 g^{tt}\w= \frac 18 \int_0^1dy \frac{(1-y)\w\eta^2e^\dd}{ay^2}\,,
\eqlabel{chargedensitytoy}
\end{equation}
where the overall normalization  in \eqref{bscurrent} is fixed to agree with the $\hM-{\rm vs.}-\hQ$ curve for the ground state boson stars in supergravity in the limit 
$\hQ\to 0$, see section \ref{full}. Since $\hQ\propto (\eta^b_3)^2+\calo((\eta^b_3)^4)$, we can use $\eta^b_3$ as a proxy for the boson star charge 
$\hQ$; then, for a fixed $\eta^b_3$, the asymptotics of the solution \eqref{bstoyuv} and \eqref{bstoyir} are characterized by 4 parameters 
\begin{equation}
\{\w\,,\ a^b_2\,,\ \eta^h_0\,,\ d^h_0\}\,,
\eqlabel{partoy}
\end{equation}
which is the overall order of the ODE system \eqref{bstoye1}-\eqref{bstoye3}. The mass of a boson star is given by \eqref{mq} and  \eqref{mqred}
\begin{equation}
\hM=-4 a^b_2\,.
\eqlabel{toymass1}
\end{equation}

It is possible to construct analytic solutions to \eqref{bstoye1}-\eqref{bstoye3} perturbatively in the amplitude of $\eta$:
\begin{equation}
\begin{split}
&\eta(y)=\sum_{n=0}^{\infty} \l^{2n+1}\ \eta_{[2n+1]}(y)\,,\qquad a=1+\sum_{n=1}^{\infty} \l^{2n}\ a_{[2n]}(y)\,,\qquad \dd=\sum_{n=1}^{\infty} \l^{2n}\ \dd_{[2n]}(y) \\
&\w=\sum_{n=0}^{\infty} \l^{2n}\w_{[2n]}\,,\qquad \w_{[2n]}={\rm const}\,.
\end{split}
\eqlabel{perttoy}
\end{equation}
To leading order we find:
\begin{equation}
\begin{split}
&\eta_{[1]}=\eta_{[1,j]}=(-1)^j\ _2F_1\left(-j,3+j;2,1-y\right)\ y^{3/2}\,,\qquad \w_{[0]}=\w_{[0,j]}=3+2 j\,,
\end{split}
\eqlabel{toypertexp}
\end{equation}
where $j=0,1,\cdots$ is an index labeling different branches of boson star solutions.
Furthermore, we find
\begin{equation}
\begin{split}
&\hQ=\hQ_{j}=\frac{\l^2}{8(j+1)(j+2)}+\calo(\l^4)\,,\\
&\hM=\hM_{j}=\frac{2(3+2j)\l^2}{3(j+1)(j+2)}+\calo(\l^4)=16\ \frac{\w_{j}}{3}\ Q_j+\calo(Q_j^2)\,. 
\end{split}
\eqlabel{mqtoy2}
\end{equation}
Note that for a fixed $\hQ$, higher $j$-level boson stars are more massive:
\begin{equation}
\hM_{j+k}(\hQ) > \hM_j(\hQ)\,,\qquad k=1,2\cdots\,.
\eqlabel{moremassive}
\end{equation}

\begin{figure}[t]
\begin{center}
\psfrag{q}{{$\hQ$}}
\psfrag{m}{{$\hM$}}
\psfrag{w}{{$\omega_{j=0}$}}
\includegraphics[width=3.0in]{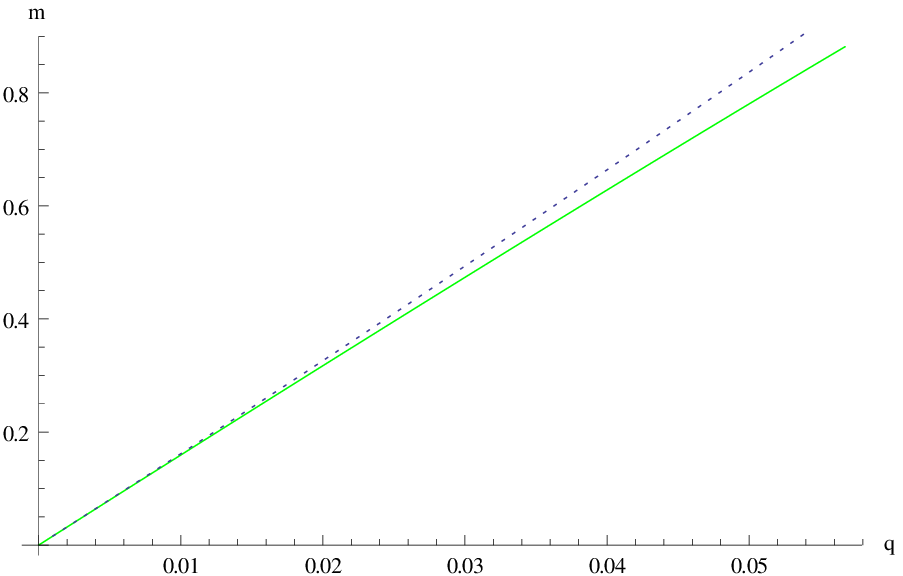}
\includegraphics[width=3.0in]{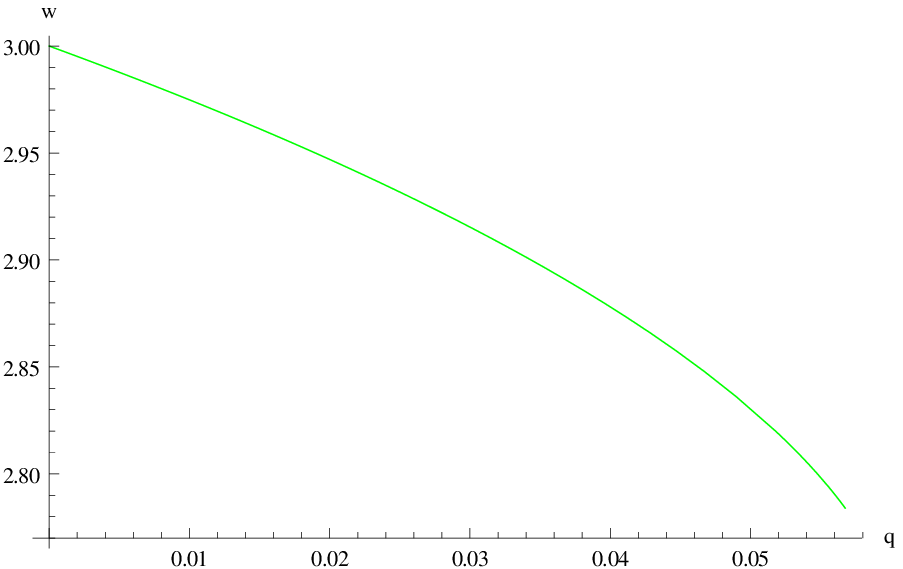}
\end{center}
  \caption{Left panel: The mass of the $j=0$ phenomenological boson star (solid green line), and the mass 
of the lightest RN black hole with the same charge (the dotted blue line). Right panel: frequency of the $j=0$ boson star.} \label{figure3}
\end{figure}

Figure \ref{figure3} represent numerical results for the $j=0$ phenomenological boson stars, fully nonlinear in $\hQ$.
Left panel: the solid green line represents the mass of the ground state boson stars as a function of the charge $\hQ$.
For reference, the dotted blue line is the mass of the lightest RN black hole \eqref{mmin} at the same charge. 
The right panel represents the ground state frequency of a boson star as a function of $\hQ$.

\begin{figure}[t]
\begin{center}
\psfrag{x}{{$\ln(\eta^b_{3,critical}-\eta^b_3)$}}
\psfrag{y}{{$\{\frac\w3,a^b_2\}$}}
\psfrag{z}{{$\{\eta^h_0,d^h_0\}$}}
\includegraphics[width=2.5in]{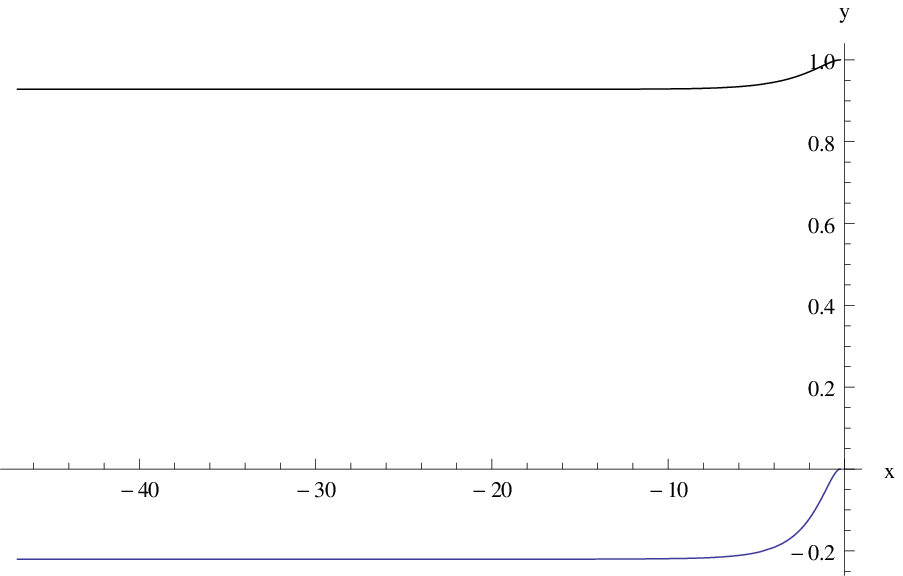}
\includegraphics[width=2.5in]{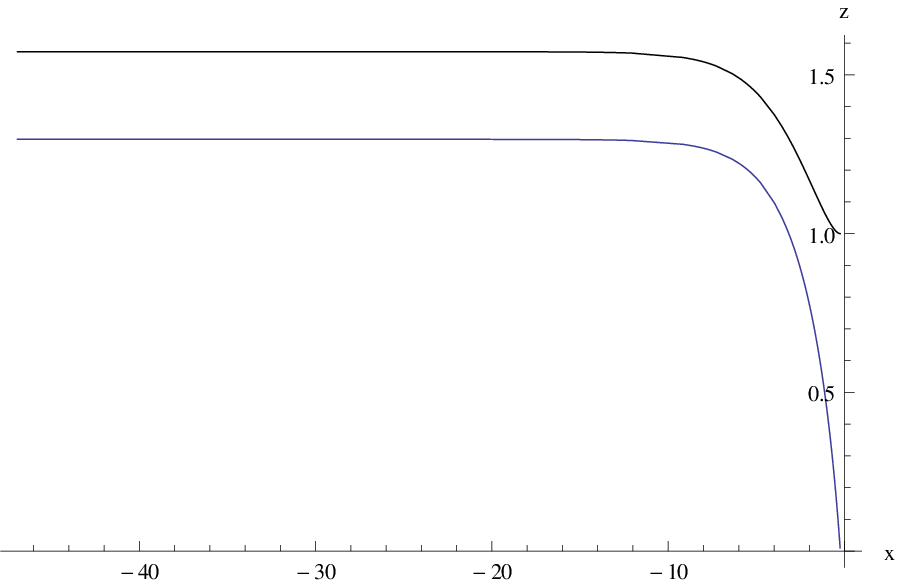}
\end{center}
  \caption{Dependence of the $j=0$ phenomenological boson star parameters \eqref{partoy} on $\eta^b_3$. Left panel: $a^b_2$ (blue curve), $\frac\w3$ 
(black curve); right panel: $\eta^h_0$ (blue curve), $d^h_0$ (black curve).  For $\eta^c_{3,critical}$ we used the numerical upper limit as in \eqref{etac}. } \label{figure3a}
\end{figure}

Unlike black holes, boson stars do not exist\footnote{Similar phenomenon was observed in \cite{Astefanesei:2003qy,Buchel:2013uba,Buchel:2015sma}.} 
beyond some critical value of $\eta^b_3$ (or corresponding the charge $\hQ$),
\begin{equation}
|\eta^b_3|\le |\eta^c_{3,critical}|\ <\ 0.7887057630169999778594689006317396074408\,.
\eqlabel{etac}
\end{equation} 
As Figure \ref{figure3a} demonstrates, there is no indication in the numerical results for \eqref{partoy}  
why there is a bound for  $\eta^b_3$. Left panel presents results for $a^b_2$ (blue curve) and $\frac \w3$ (black curve)
as a function of $\eta^b_3$. For $\eta^c_{3,critical}$ we used the numerical upper limit as in \eqref{etac}. 
Right panel presents analogous results for $\eta^h_0$ (blue curve) and $d^h_0$ (black curve). 
It would be interesting to understand the physical reason for the existence of  $\eta^b_{3,critical}$ ---
previous linearized stability analysis of the phenomenological boson stars in \cite{Astefanesei:2003qy,Buchel:2015sma} found  instabilities 
{\it prior to } reaching the analog of $\eta^b_{3,critical}$.

\section{Boson stars in supergravity}\label{full}

To find boson stars in GHPT model \eqref{ghptaction} we use metric ansatz \eqref{metric5d} 
along with 
\begin{equation}
A=\phi(y)\ d(t)\,,\qquad c\equiv \cosh(\eta)=c(y)\,,\qquad \theta=\w t\,.
\eqlabel{fullextra}
\end{equation}
The corresponding equations of motion are similar to \eqref{bh1}-\eqref{bh4} 
with the replacements 
\begin{equation}
q^2\phi^2\ \to\ (q \phi -\w)^2\,,
\eqlabel{replace1}
\end{equation}
in all terms except for the last one in \eqref{bh4} where we replace 
\begin{equation}
q^2\phi\ \to\ q(q \phi -\w)^2\,.
\eqlabel{replace2}
\end{equation}
Type IIB supergravity embedding of GHPT model sets $q=2$.

Asymptotic data for  GHPT boson stars take form:
\nxt at the $AdS_5$ boundary, $y\to 0_+$, 
\begin{equation}
\begin{split}
&a=1+a^b_2\ y^2+\left(a^b_2+\frac49 (\phi^b_1)^2+c^b_3\right)\ y^3+\calo(y^4)\,,\qquad \dd=\frac12 c^b_3\ y^3+\calo(y^4)\,,\\
&c=1+c^b_3\ y^3+\calo(y^4)\,,\qquad \phi=\phi^b_1\ y+\phi^b_1\ y^2+\left(\phi^b_1-\frac{1}{16}q \w c^b_3 \right)\ y^3+\calo(y^4)\,;
\end{split}
\eqlabel{bsfulluv}
\end{equation}
\nxt at the origin, $z\equiv 1-y\to 0_+$,
\begin{equation}
\begin{split}
&a=1-\biggl(
\frac{1}{24} (d^h_0)^2 (\phi^h_0)^2 ((c^h_0)^2-1) q^2-\frac{1}{12} (d^h_0)^2 \w \phi^h_0 ((c^h_0)^2-1) q+\frac{1}{24} (c^h_0-1) 
(c^h_0 (d^h_0)^2 \w^2\\
&+(d^h_0)^2 \w^2+3 c^h_0-9)
\biggr)\ z+\calo(z^2)\,,\\
&\dd=\ln d^h_0-\frac{1}{12} (d^h_0)^2 (\phi^h_0 q-\w)^2 ((c^h_0)^2-1)\ z+\calo(z^2)\,,\\
&\phi=\phi^h_0+\frac{3}{32} q ((c^h_0)^2-1) (\phi^h_0 q-\w)\ z+\calo(z^2)\,,\\
&c=c^h_0-\frac18 ((c^h_0)^2-1) (c^h_0 (d^h_0)^2 (\phi^h_0 q-\w)^2-3 c^h_0+6)\ z+\calo(z^2)\,.
\end{split}
\eqlabel{bsfullir}
\end{equation}
The mass and the charge of the boson stars are given by \eqref{mqhair}. Note that for a fixed $\hQ$ (equivalently $\phi^b_1$), 
the asymptotics of the solution are characterized by 6 parameters, 
\begin{equation}
\{\w\,,\ a^b_2\,,\ c^b_3\,,\ \phi^h_0\,,\ c^h_0\,,\ d^h_0\}\,,
\eqlabel{fullpar1}
\end{equation}
which is  the overall order of the coupled ODE system of the equations of motion.

As in section \ref{toy}, we can construct boson stars perturbatively in the $\eta(y)$ amplitude analytically. We have done it 
for the $j=0$ boson stars in GHPT model to order $\calo(\l^{30})$ inclusive, where $\l$ is fixed as 
\begin{equation}
c^b_3\equiv \frac 12\l^2\,.
\eqlabel{dixl}
\end{equation}  
We find:
\begin{equation}
\begin{split}
&\hQ\bigg|_{j=0}=-\frac 13 \phi^b_1\\
&= \frac{1}{16} \l^2+\frac{3}{256} \l^4+\frac{29}{5120} \l^6+\frac{8441}{2293760} \l^8+\frac{1903}{688128} \l^{10}+\frac{68603543}{30277632000} \l^{12}
\\
&+\frac{2357849941}{1199570944000} \l^{14}+\frac{140271611444141}{78998944088064000} \l^{16}+\frac{148011392983393047}{89532136633139200000} \l^{18}
\\
&+\frac{28945266865118145121}{18371994437120163840000} \l^{20}
+\frac{45418739444730506807}{29692112221608345600000} \l^{22}\\
&+\frac{619893865521183402070738489}{411131694835982011465728000000} \l^{24}
+\frac{5303518796142734229022251479}{3523985955736988669706240000000} \l^{26}
\\&+\frac{9769477519044177078641271669913}{6435072951483980556753960960000000} \l^{28}
\\&+\frac{165805336474260222939581033601087896467}{107292400324955840088129241350144000000000} \l^{30}+\calo(\l^{32})\,,
\end{split}
\eqlabel{hqfullbs}
\end{equation}
and, remarkably,  
\begin{equation}
\w\bigg|_{j=0}=3+\calo(\l^{32})\,,\qquad \hM\bigg|_{j=0}=16\ \hQ + \calo(\l^{32})\,.
\eqlabel{hqfullbs1}
\end{equation}
Turns out that relations \eqref{hqfullbs1} hold only for the ground state, \ie $j=0$, and only for the supergravity value  $q=2$:
\begin{equation}
\begin{split}
&\w\bigg|_{j=1}=5+\left(-\frac{62}{105}+\frac{1}{28}q^2\right)\ \l^2
+\biggl(-\frac{38170019869}{63567504000}-\frac{9328183}{9040711680}q^4\,,\\
&+\frac{1871544527}{33902668800}q^2\biggr)\ \l^4+\calo(\l^6)\\
&\hM\bigg|_{j=1}=\left(\frac{160}{3q}+\frac{4(15q^2-248)}{315q}\ \lambda^2\right)\ \hQ+\calo(\l^4)\,.
\end{split}
\eqlabel{fullratios}
\end{equation}
Precision fully nonlinear numerics confirmed that  identities \eqref{hqfullbs1} are in fact exact.

\begin{figure}[t]
\begin{center}
\psfrag{n}{{$n$}}
\psfrag{c}{{$\ln |c^b_{3,n}|$}}
\includegraphics[width=4.0in]{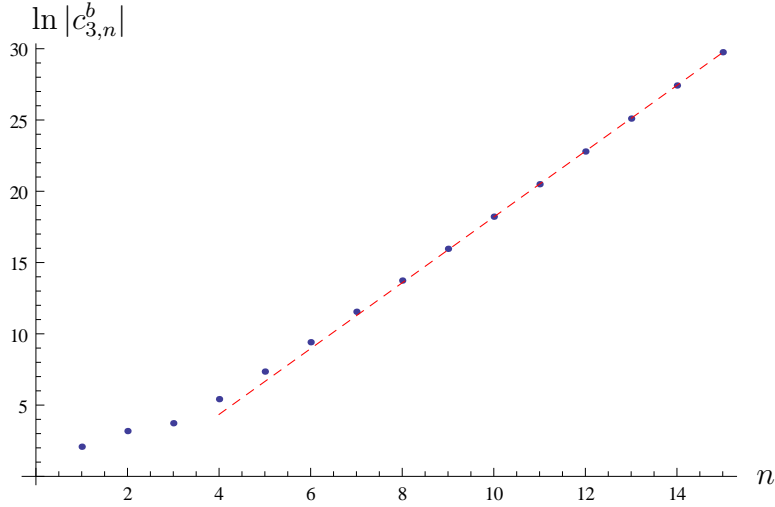}
\end{center}
  \caption{Perturbative expansion of $c^b_3$ boson star parameter has a finite radius of convergence in $\hQ$. 
The data points represent the first few expansion coefficients, see \eqref{critfullund}, and the red dashed line 
is the linear fit to the last five data points, see \eqref{fit}.
} \label{figure4}
\end{figure}

As in case of the phenomenological boson stars discussed in section \ref{toy}, here there is also a bound on $c^b_3$; we find
\begin{equation}
c^b_3\le c^b_{3,critical} < 0.43779161842835156074386507215849905168975\,.
\eqlabel{cb3crit}
\end{equation}
Unlike phenomenological boson stars discussed in section \ref{toy}, here we can understand (at least at a technical level) 
the origin of $c^b_{3,critical}$. We can invert \eqref{hqfullbs},
\begin{equation}
\begin{split}
&c^b_3=\sum_{n=1}^{\infty} c^b_{3,n}\ \hQ^n\\
=&8 \hQ-24 \hQ^2-\frac{208}{5} \hQ^3-\frac{7888}{35} \hQ^4-\frac{818176}{525} \hQ^5-\frac{352876624}{28875} \hQ^6-\frac{272007245344}{2627625} \hQ^7\\
&-\frac{84964430758448}{91966875} \hQ^8-\frac{18253061040643088}{2131959375} \hQ^9-\frac{109122556608102498928}{1336738528125} \hQ^{10}\\
&-\frac{9101984161925699661824}{11436540740625} \hQ^{11}-\frac{553801482686949813868971824}{70110393409546875} \hQ^{12}\\
&-\frac{61208627756373246966742061152}{769393278325546875} \hQ^{13}-\frac{5249319592830812634726466094176}{6471955223561953125} \hQ^{14}\\
&-\frac{97973604596679544890714472505785440016}{11725726675168957623046875} \hQ^{15}+\calo(\hQ^{16})\,.
\end{split}
\eqlabel{critfullund}
\end{equation}
Computed coefficients $c^b_{3,n}$ are presented for $n=1,\cdots 15$ as blue dots in Figure \ref{figure4}. The dashed red line is the linear  fit to the 
last five data points:
\begin{equation}
\ln|c^{b}_{3,n}|\bigg|_{red,dashed}=-4.88369 + 2.30773\ n \,.
\eqlabel{fit}
\end{equation}   
The extracted asymptotic behavior of the coefficients $c^{b}_{3,n}$ suggests that the radius of convergence of the perturbative expansion  
\eqref{critfullund} is 
\begin{equation}
\hQ \lesssim e^{-2.30773}=0.0994864\qquad \Longrightarrow\qquad  c^b_3\lesssim 0.395827\,,
\end{equation}
corresponding to $c^b_{3}$ bound close to \eqref{cb3crit}.

\begin{figure}[t]
\begin{center}
\psfrag{q}{{$\hQ$}}
\psfrag{m}{{$\hM$}}
\includegraphics[width=5.0in]{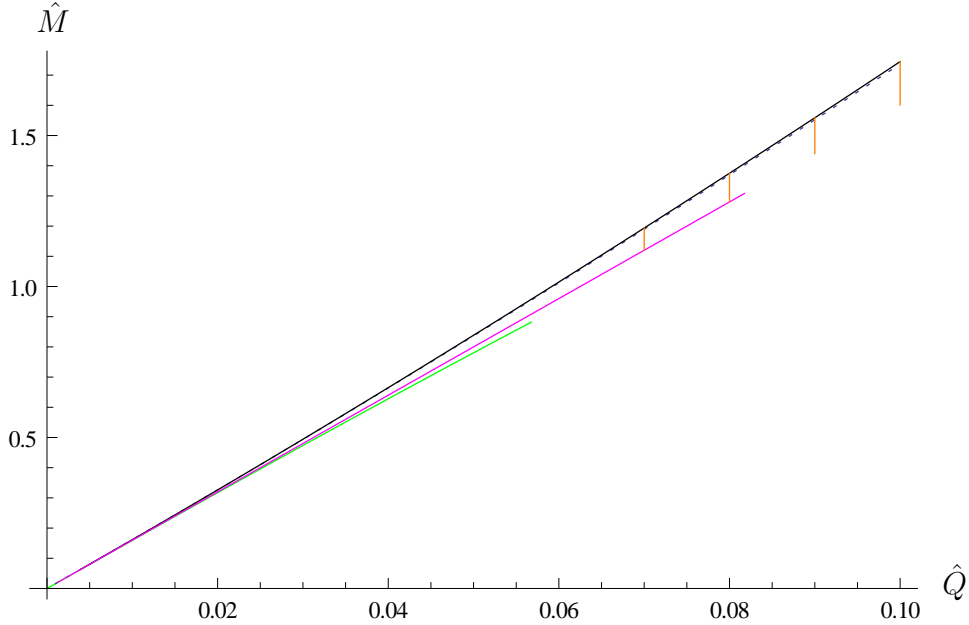}
\end{center}
  \caption{Charged states in GHPT model: the solid black curve represents RN black holes at the threshold of 
$\Psi$-condensation instability; dotted blue curve is the smallest mass RN black hole for a given charge $\hQ$;
orange curves are RN-$\Psi$ black holes for select values of $\hQ$; the magenta line represents the ground state
GHPT boson stars. The solid green curve represents ground state phenomenological boson stars.} \label{figure5}
\end{figure}

Figure \ref{figure5} collects the mass-charge dependence of various states in GHPT model. 
The solid black curve is the onset of the $\Psi$-scalar instability for RN black holes; the 
dotted blue curve is the smallest mass of the RN black hole for a  given $\hQ$. The orange 
curves represent the RN-$\Psi$ black holes (see section \ref{bhbh}) for select values of $\hQ$. 
They terminate (at low mass) as the RN-$\Psi$ black hole becomes vanishingly small. Our numerical results 
indicate that they terminate slightly above the magenta line, which represents the 
mass-charge relation for the ground state boson stars in GHPT:
\begin{center}
  \begin{tabular}{ | c | c | c|  }
 \hline
 $\hQ$ & min $\hM$  [RN-$\Psi$ black hole] & $\hM$[ boson star]\\   \hline
 $0.07$ &1.120470(7)   & 1.12	 \\  
\hline
 $0.08$ & 1.280128(6)  & 1.28	 \\  
\hline
  \end{tabular}
\end{center}
Numerically, 
we can not access zero size RN-$\Psi$ black holes --- it is possible that the precise $y_0\to 0$ limit 
of RN-$\Psi$ black holes saturates the $\hM=16 \hQ$ relation of the $j=0$ boson stars. 
Finally, the solid green curve is the mass versus charge relation for the phenomenological 
$j=0$ boson stars discussed in section \ref{toy}.

\section{Conclusion}\label{conclude}

In this paper we studied boson stars --- gravitational soliton-like configurations supported by a complex scalar field 
--- in type IIB supergravity. More precisely, we focused on GHPT model \cite{Gubser:2009qm},
representing the holographic dual to a class of strongly coupled $\caln=1$ quiver conformal gauge theories with 
a dimension $\Delta=3$ chiral primary operator $\calo_3$. Earlier work \cite{Buchel:2013uba,Buchel:2015sma} indicated
that boson stars (and boson-star-like configurations ) play an important role in low-energy dynamics of holographic gauge theories,
in particular regarding the question of equilibration. These earlier work (also \cite{Astefanesei:2003qy}), however,
were restricted to boson stars with global $U(1)$ symmetry. In gravitational holographic duals to gauge theories the global symmetries
must be gauged. The role of the corresponding local symmetry is played by  $R$-symmetry in GHPT holography.  
  
We studied in details the  spectrum of charged states in GHPT plasma confined on $S^3$ in a microcanonical ensemble. 
First, there are thermal states with $\langle\calo_3\rangle =0$. These are represented by
Reissner-Nordstrom (RN)
black holes in global $AdS_5$. While important at high energy, these states are not expected to dominated the low-energy 
dynamics. On one hand, general arguments \cite{Horowitz:1999uv} suggest that they will be unstable\footnote{The instability was 
explicitly demonstrated only for neutral thermal states so far \cite{Hubeny:2002xn,Dias:2015pda,Buchel:2015gxa,Buchel:2015pla}.} 
with respect to localization on a transverse compact space, spontaneously breaking the $R$-symmetry. In GHPT model there is 
a more mundane instability associated with  developing the condensate $\langle\calo_3\rangle \ne 0$ at sufficiently low 
energy for a fixed charge. These new states are thermal, and are represented by RN-$\Psi$ black holes. RN-$\Psi$ black holes
reach below the mass threshold for the existence of RN black holes, almost (but not quite according to our numerics) reaching the 
ground state branch of the boson stars. The ground state GHPT boson stars represent remarkable configurations: unlike their counterparts in 
phenomenological models, their frequency is constant; furthermore, they saturate a BPS-like relation between the mass and the charge. 
However, these boson stars exist only below certain charge threshold.
  
We left many questions unanswered. 
It would be interesting to explicitly establish the GL instability of RN black holes following \cite{Buchel:2015pla}.
Likewise, it is important to establish the stability of GHPT boson stars, extending the work \cite{Buchel:2015sma}.
Is there a gap in the spectrum of RN-$\Psi$ black holes and the boson stars? Is it possible to construct 
charged initial conditions with energy less than that of a GHPT boson star?    
We hope to report on these questions in the future.

%%%%%%%%%%%%%%%%%%%%%%%%%%%%%%%%%%%%%%%%%%%%%%%%%%%%%%%%%%%%%%%%%%%%

\section*{Acknowledgments}
I would like to thank Larry Yaffe for valuable discussions. 
Research at Perimeter
Institute is supported by the Government of Canada through Industry
Canada and by the Province of Ontario through the Ministry of
Research \& Innovation. This work was further supported by
NSERC through the Discovery Grants program.


\begin{thebibliography}{99}


%\cite{Liebling:2012fv}
\bibitem{Liebling:2012fv} 
  S.~L.~Liebling and C.~Palenzuela,
  ``Dynamical Boson Stars,''
  Living Rev.\ Rel.\  {\bf 15}, 6 (2012)
  [arXiv:1202.5809 [gr-qc]].
  %%CITATION = ARXIV:1202.5809;%%
  %59 citations counted in INSPIRE as of 24 Oct 2015


%\cite{Astefanesei:2003qy}
\bibitem{Astefanesei:2003qy} 
  D.~Astefanesei and E.~Radu,
  ``Boson stars with negative cosmological constant,''
  Nucl.\ Phys.\ B {\bf 665}, 594 (2003)
  [gr-qc/0309131].
  %%CITATION = GR-QC/0309131;%%
  %47 citations counted in INSPIRE as of 24 Oct 2015

%\cite{Bizon:2011gg}
\bibitem{Bizon:2011gg} 
  P.~Bizon and A.~Rostworowski,
  ``On weakly turbulent instability of anti-de Sitter space,''
  Phys.\ Rev.\ Lett.\  {\bf 107}, 031102 (2011)
  [arXiv:1104.3702 [gr-qc]].
  %%CITATION = ARXIV:1104.3702;%%
  %176 citations counted in INSPIRE as of 24 Oct 2015

%\cite{Buchel:2013uba}
\bibitem{Buchel:2013uba} 
  A.~Buchel, S.~L.~Liebling and L.~Lehner,
  ``Boson stars in AdS spacetime,''
  Phys.\ Rev.\ D {\bf 87}, no. 12, 123006 (2013)
  [arXiv:1304.4166 [gr-qc]].
  %%CITATION = ARXIV:1304.4166;%%
  %64 citations counted in INSPIRE as of 24 Oct 2015

%\cite{Aharony:1999ti}
\bibitem{Aharony:1999ti} 
  O.~Aharony, S.~S.~Gubser, J.~M.~Maldacena, H.~Ooguri and Y.~Oz,
  ``Large N field theories, string theory and gravity,''
  Phys.\ Rept.\  {\bf 323}, 183 (2000)
  [hep-th/9905111].
  %%CITATION = HEP-TH/9905111;%%
  %3571 citations counted in INSPIRE as of 24 Oct 2015

%\cite{Horowitz:1999uv}
\bibitem{Horowitz:1999uv} 
  G.~T.~Horowitz,
  ``Comments on black holes in string theory,''
  Class.\ Quant.\ Grav.\  {\bf 17}, 1107 (2000)
  [hep-th/9910082].
  %%CITATION = HEP-TH/9910082;%%
  %45 citations counted in INSPIRE as of 12 Jan 2015


%\cite{Gregory:1993vy}
\bibitem{Gregory:1993vy} 
  R.~Gregory and R.~Laflamme,
  ``Black strings and p-branes are unstable,''
  Phys.\ Rev.\ Lett.\  {\bf 70}, 2837 (1993)
  [hep-th/9301052].
  %%CITATION = HEP-TH/9301052;%%
  %671 citations counted in INSPIRE as of 02 Feb 2015



%\cite{Hubeny:2002xn}
\bibitem{Hubeny:2002xn} 
  V.~E.~Hubeny and M.~Rangamani,
  ``Unstable horizons,''
  JHEP {\bf 0205}, 027 (2002)
  [hep-th/0202189].
  %%CITATION = HEP-TH/0202189;%%
  %54 citations counted in INSPIRE as of 28 gen 2015

%\cite{Dias:2015pda}
\bibitem{Dias:2015pda} 
  O.~J.~C.~Dias, J.~E.~Santos and B.~Way,
  ``Lumpy AdS$_{5}$× S$^{5}$ black holes and black belts,''
  JHEP {\bf 1504}, 060 (2015)
  [arXiv:1501.06574 [hep-th]].
  %%CITATION = ARXIV:1501.06574;%%
  %3 citations counted in INSPIRE as of 23 sept. 2015

%\cite{Buchel:2015gxa}
\bibitem{Buchel:2015gxa} 
  A.~Buchel and L.~Lehner,
  ``Small black holes in $AdS_5\times S^5$,''
  Class.\ Quant.\ Grav.\  {\bf 32}, no. 14, 145003 (2015)
  [arXiv:1502.01574 [hep-th]].
  %%CITATION = ARXIV:1502.01574;%%
  %3 citations counted in INSPIRE as of 24 Oct 2015



%\cite{Buchel:2015pla}
\bibitem{Buchel:2015pla} 
  A.~Buchel,
  ``Universality of small black hole instability in AdS/CFT,''
  arXiv:1509.07780 [hep-th].
  %%CITATION = ARXIV:1509.07780;%%

%\cite{Buchel:2015sma}
\bibitem{Buchel:2015sma} 
  A.~Buchel and M.~Buchel,
  ``On stability of nonthermal states in strongly coupled gauge theories,''
  arXiv:1509.00774 [hep-th].
  %%CITATION = ARXIV:1509.00774;%%

%\cite{Gubser:2009qm}
\bibitem{Gubser:2009qm} 
  S.~S.~Gubser, C.~P.~Herzog, S.~S.~Pufu and T.~Tesileanu,
  ``Superconductors from Superstrings,''
  Phys.\ Rev.\ Lett.\  {\bf 103}, 141601 (2009)
  [arXiv:0907.3510 [hep-th]].
  %%CITATION = ARXIV:0907.3510;%%
  %186 citations counted in INSPIRE as of 24 Oct 2015

%\cite{Kehagias:1998gn}
\bibitem{Kehagias:1998gn} 
  A.~Kehagias,
  ``New type IIB vacua and their F theory interpretation,''
  Phys.\ Lett.\ B {\bf 435}, 337 (1998)
  [hep-th/9805131].
  %%CITATION = HEP-TH/9805131;%%
  %149 citations counted in INSPIRE as of 24 Oct 2015

%\cite{Klebanov:1998hh}
\bibitem{Klebanov:1998hh} 
  I.~R.~Klebanov and E.~Witten,
  ``Superconformal field theory on three-branes at a Calabi-Yau singularity,''
  Nucl.\ Phys.\ B {\bf 536}, 199 (1998)
  [hep-th/9807080].
  %%CITATION = HEP-TH/9807080;%%
  %865 citations counted in INSPIRE as of 24 Oct 2015

%\cite{Acharya:1998db}
\bibitem{Acharya:1998db} 
  B.~S.~Acharya, J.~M.~Figueroa-O'Farrill, C.~M.~Hull and B.~J.~Spence,
  ``Branes at conical singularities and holography,''
  Adv.\ Theor.\ Math.\ Phys.\  {\bf 2}, 1249 (1999)
  [hep-th/9808014].
  %%CITATION = HEP-TH/9808014;%%
  %205 citations counted in INSPIRE as of 24 Oct 2015


%\cite{Morrison:1998cs}
\bibitem{Morrison:1998cs} 
  D.~R.~Morrison and M.~R.~Plesser,
  ``Nonspherical horizons. 1.,''
  Adv.\ Theor.\ Math.\ Phys.\  {\bf 3}, 1 (1999)
  [hep-th/9810201].
  %%CITATION = HEP-TH/9810201;%%
  %371 citations counted in INSPIRE as of 24 Oct 2015

%\cite{Buchel:2008ae}
\bibitem{Buchel:2008ae} 
  A.~Buchel, R.~C.~Myers, M.~F.~Paulos and A.~Sinha,
  ``Universal holographic hydrodynamics at finite coupling,''
  Phys.\ Lett.\ B {\bf 669}, 364 (2008)
  [arXiv:0808.1837 [hep-th]].
  %%CITATION = ARXIV:0808.1837;%%
  %90 citations counted in INSPIRE as of 25 Oct 2015


%\cite{Buchel:2006gb}
\bibitem{Buchel:2006gb} 
  A.~Buchel and J.~T.~Liu,
  ``Gauged supergravity from type IIB string theory on Y**p,q manifolds,''
  Nucl.\ Phys.\ B {\bf 771}, 93 (2007)
  [hep-th/0608002].
  %%CITATION = HEP-TH/0608002;%%
  %70 citations counted in INSPIRE as of 25 Oct 2015



%\cite{Aharony:2007vg}
\bibitem{Aharony:2007vg} 
  O.~Aharony, A.~Buchel and P.~Kerner,
  ``The Black hole in the throat: Thermodynamics of strongly coupled cascading gauge theories,''
  Phys.\ Rev.\ D {\bf 76}, 086005 (2007)
  [arXiv:0706.1768 [hep-th]].
  %%CITATION = ARXIV:0706.1768;%%
  %53 citations counted in INSPIRE as of 28 Aug 2015



\end{thebibliography}
\end{document}